\documentclass[prd,superscriptaddress,showpacs,nofootinbib,amsmae fi,amssymb,aps,11pt]{revtex4}

\usepackage{bm}
\usepackage{amsfonts}
\usepackage{latexsym}
\usepackage[latin1]{inputenc}
\usepackage{graphicx}
\usepackage{amsmath}
\usepackage{palatino}
\usepackage{mathpazo}
\usepackage{booktabs}
\usepackage{dcolumn}

\def\jnl@style{\it}
\def\aaref@jnl#1{{\jnl@style#1}}

\def\aaref@jnl#1{{\jnl@style#1}}

\def\aj{\aaref@jnl{AJ}}                   
\def\apj{\aaref@jnl{ApJ}}                 
\def\apjl{\aaref@jnl{ApJ}}                
\def\apjs{\aaref@jnl{ApJS}}               
\def\apss{\aaref@jnl{Ap\&SS}}             
\def\aap{\aaref@jnl{A\&A}}                
\def\aapr{\aaref@jnl{A\&A~Rev.}}          
\def\aaps{\aaref@jnl{A\&AS}}              
\def\mnras{\aaref@jnl{Mon.~Not.~Roy.~Astron.~Soc.}}             
\def\prd{\aaref@jnl{Phys.~Rev.~D}}        
\def\prc{\aaref@jnl{Phys.~Rev.~C}}  
\def\prl{\aaref@jnl{Phys.~Rev.~Lett.}}    
\def\qjras{\aaref@jnl{QJRAS}}             
\def\skytel{\aaref@jnl{S\&T}}             
\def\ssr{\aaref@jnl{Space~Sci.~Rev.}}     
\def\zap{\aaref@jnl{ZAp}}                 
\def\nat{\aaref@jnl{Nature}}              
\def\aplett{\aaref@jnl{Astrophys.~Lett.}} 
\def\apspr{\aaref@jnl{Astrophys.~Space~Phys.~Res.}} 
\def\physrep{\aaref@jnl{Phys.~Rep.}}      
\def\physscr{\aaref@jnl{Phys.~Scr}}       
\def\commat{\aaref@jnl{Comm.~Math.~Phys.}}              
\def\science{\aaref@jnl{Science}}               
\def\cqg{\aaref@jnl{Classical Quant.~Grav.}}            
\def\jpcs{\aaref@jnl{JPCS}}                                     
\def\ijmpd{\aaref@jnl{Int.~J.~Mod.~Phys.~D}}                    
\def\grg{\aaref@jnl{Gen.~Relat.~Gravit.}}               
\def\rpp{\aaref@jnl{Rep.~Prog.~Phys.}}          
\def\npa{\aaref@jnl{Nucl.~Phys.~A}}        
\def\lrr{\aaref@jnl{Living Rev.~Rel.}}                   
\def\jcap{\aaref@jnl{J.~Cosmology Astropart.~Phys.}}    
\def\rmp{\aaref@jnl{Rev.~Mod.~Phys.}}   

\allowdisplaybreaks[1]

\begin{document}

\title{Rapidly rotating neutron stars with a massive scalar field -- structure and universal relations}

\author{Daniela D. Doneva}
\email{daniela.doneva@uni-tuebingen.de}
\affiliation{Theoretical Astrophysics, Eberhard Karls University of T\"ubingen, T\"ubingen 72076, Germany}
\affiliation{INRNE - Bulgarian Academy of Sciences, 1784  Sofia, Bulgaria}

\author{Stoytcho S. Yazadjiev}
\affiliation{Theoretical Astrophysics, Eberhard Karls University of T\"ubingen, T\"ubingen 72076, Germany}
\affiliation{Department of Theoretical Physics, Faculty of Physics, Sofia University, Sofia 1164, Bulgaria}


\begin{abstract}
We construct rapidly rotating neutron star models in scalar-tensor theories with a massive scalar field. The fact that the scalar field has nonzero mass leads to very interesting results since the allowed range of values of the coupling parameters is significantly broadened. These deviations from pure general relativity can be very large for values of the parameters that are in agreement with the observations. The rapid rotation can  magnify the differences several times compared to the static case. The universal relations between the normalized moment of inertia and quadrupole moment are also investigated both for the slowly and rapidly rotating cases. The results show that these relations are still EOS independent up to a large extend and the deviations from pure general relativity can be large. This places the massive scalar-tensor theories amongst the few alternative theories of gravity that can be tested via the universal $I$-Love-$Q$ relations.
\end{abstract}


\maketitle

\section{Introduction}
In the recent years, alternative theories of gravity attracted considerable interest. Compact objects, such as neutron stars and black holes, were considered in various generalizations of Einstein's theory and astrophysical implications were studied that can impose constraints on these theories (for a review see \cite{Berti2015}). But still very few of the examined alternative theories of gravity can pass through all the observations, do not have any intrinsic problems and still lead to large deviations in the observational properties of the compact objects\cite{Damour1992}. 

In the present paper we will concentrate exactly on one such representative, namely the scalar-tensor theories with massive scalar field that attracted interest recently \cite{Perivolaropoulos2010,Hohmann2013,Schaerer2014,Jaerv2015,Alsing2012,Chen2015,Ramazanouglu2016,Yazadjiev2016}. Scalar-tensor theories in general are assumed to be one of the most natural generalizations of Einstein's theory and that is why they were extensively studied in the literature. Moreover, the $f(R)$ theories of gravity are mathematically equivalent to a particular class of scalar-tensor theories with nonzero potential for the scalar field. As far as neutron stars are concerned, a special attention was paid to a specific class of scalar-tensor theories which is indistinguishable from general relativity in the weak field limit but can lead to interesting strong field effects, such as scalarization and non-uniqueness of the solutions. Neutron stars in such theories with massless scalar field were considered for the first time in \cite{Damour1993,Damour1996}. Different aspects and astrophysical implications of these compact objects were studied in the literature mainly in the static case \cite{Sotani04,Sotani2005,DeDeo2003,Novak1998,Harada1997a,Harada1998,Sotani2012,Barausse2013,Palenzuela2014,Shibata2014,Silva2015} (see also \cite{Maselli2016,Minamitsuji2016}). However, the recent observations of neutron stars in compact binaries have limited severely the possible range of values of the coupling parameter \cite{Freire2012,Antoniadis13}. Thus, the deviations from the pure general relativistic solutions in the static case are very small and only the rapidly rotating case leaves space for larger differences, since the rapid rotation magnifies the deviations and widens the range of parameters where scalarized neutron stars (i.e. with nontrivial scalar field) exist \cite{Doneva2013}.

We can cure this problem and achieve large differences with pure Einstein's theory for values of the parameters that are in agreement with the observations if we consider nonzero mass of the scalar field. This effectively suppresses the scalar field at distances larger than the Compton wave-length. Therefore, the observations (both weak field and the binary pulsar observations) can be reconciled with the theory for a broad set ot parameters. This was indeed shown to be true in \cite{Popchev2015,Ramazanouglu2016,Yazadjiev2016} where neutron stars in massive scalar-tensor theories were studies for the first time. Our goal in the present paper is to extend these studies to the regime of rapid rotation which is supposed to magnify even further the deviations from pure general relativity, similar to the case of massless scalar field and $f(R)$ theories \cite{Doneva12,Yazadjiev2015}. Rapidly rotating neutron star models in other alternative theories of gravity were constructed in \cite{Kleihaus2014,Kleihaus2016}.

In the present paper we will concentrate also on building universal relations for neutron stars in massive scalar-tensor theories, both in the slowly rotating and the rapidly rotating case. Our motivation comes from the fact that the EOS independent relations are usually a good tool for testing the strong field regime of gravity because the large uncertainties in the nuclear matter equation of state are taken away. We will focus on a particular type of EOS universal relations that has drawn a lot of attention recently, the so-called $I$-Love-$Q$ relations that connect the normalized moment of inertia, tidal Love number and quadrupole moment \cite{Yagi2013,Yagi2013a}. As a first step we will consider the $I$-$Q$ relation that can give us important intuition about the problem. As a matter of fact the $I$-$Q$ relation is almost indistinguishable from general relativity for the majority of alternative theories of gravity \cite{Sham2014,Kleihaus2014,Pani2014,Doneva2014a,Pappas2015a,Pappas2015}. The only known exceptions are the dynamical Chern-Simons gravity \cite{Yagi2013,Yagi2013a} and the $f(R)$ theories of gravity \cite{Doneva2015}. Taking into account that the $f(R)$ theories are mathematically equivalent to a particular class of scalar-tensor theories with nonzero potential of the scalar field, we can expect large deviations in the case considered in the present paper. Other types of EOS-independent relations were also well studies in the literature (see e.g. \cite{Lattimer2001,Urbanec2013,Bauboeck2013,AlGendy2014,Andersson98a,Tsui2005,Delsate2015,Doneva2013a,Pani2015,Pappas2015d,Breu2016,Staykov2016}).

\section{Basic equations}
The scalar-tensor theory action in the Einstein frame has the following general form
\begin{eqnarray}\label{EFA}
S=\frac{1}{16\pi G} \int d^4x \sqrt{-g}\left[ R - 2
g^{\mu\nu}\partial_{\mu}\varphi \partial_{\nu}\varphi - 4 V(\varphi)
\right] + S_{\rm
matter}(A^2(\varphi)g_{\mu\nu},\chi),
\end{eqnarray}
where $R$ is the Ricci scalar curvature with respect to the Einstein frame metric $g_{\mu\nu}$, $A(\varphi)$ controls the coupling between the matter and the scalar field\footnote{An explicit coupling between the scalar field and the matter appears only in the Einstein frame formulation of the theory. In the physical Jordan frame no such coupling exists in order to fulfill the weak equivalence principle.}, and $V(\varphi)$ is the scalar field potential. For mathematical convenience we will perform our calculations in the Einstein frame and later transform the relevant quantities in the physical Jordan frame. The Jordan frame metric ${\tilde g}_{\mu\nu}$ and the gravitational scalar $\Phi$ are  given respectively by ${\tilde g}_{\mu\nu}= A^2(\varphi)g_{\mu\nu}$ and $\Phi=A^{-2}(\varphi)$. 

The choice of $V(\varphi)$ and $A(\varphi)$ completely specifies the scalar-tensor theory and we will work with 
\begin{equation}\label{eq:Aphi}
A(\varphi)=e^{\frac{1}{2}\beta\varphi^2},
\end{equation}
where $\beta$ is a constant and
\begin{equation}
V(\varphi)=\frac{1}{2}m^2_{\varphi}\varphi^2.
\end{equation}
This choice of the scalar-field potential yields a mass of the scalar field $m_\varphi$. The particular form of the coupling function \eqref{eq:Aphi} is interesting because in this case the scalar-tensor theory is indistinguishable from general relativity in the weak field regime but for strong fields nonlinear effects can develop such as the so-called spontaneous scalarization \cite{Damour1992,Damour1996,Doneva2013}. Since our main goal in the present paper is to demonstrate the effect of rapid rotation and to study the deviations from the static case, we decided  to limit our studies only to the case given by eq. \eqref{eq:Aphi}. Moreover, other choices of $A(\varphi)$, such as the Brans-Dicke case with $A(\varphi)=e^{\alpha_0 \varphi}$, lead to qualitatively similar results at least in the static and the slowly rotating cases \cite{Yazadjiev2016}.

The field equations can be derived from the action (\ref{EFA}) and have the following form:
\begin{eqnarray}
&&R_{\mu\nu} - \frac{1}{2}g_{\mu\nu}R= 8\pi G T_{\mu\nu} + 2\nabla_{\mu}\varphi\nabla_{\nu}\varphi - g_{\mu\nu} g^{\alpha\beta}\nabla_{\alpha}\varphi\nabla_{\beta}\varphi
- 2 V(\varphi) g_{\mu\nu}, \label{eq:FieldEq1} \\
&&\nabla_{\mu}\nabla^{\mu}\varphi = -4\pi G \alpha(\varphi) T + \frac{dV(\varphi)}{d\varphi}, \label{eq:FieldEq2}
\end{eqnarray}
where $\nabla_{\mu}$  is the covariant derivative with respect to the metric $g_{\mu\nu}$, $\alpha(\varphi)$ is the coupling function defined by $\alpha(\varphi)=\frac{d\ln A(\varphi)}{d\varphi}$ and $T_{\mu\nu}$ is the Einstein frame energy-momentum tensor connected to the Jordan frame one ${\tilde T}_{\mu\nu}$ via the relation $T_{\mu\nu}=A^2(\varphi){\tilde T}_{\mu\nu}$. For a perfect fluid the relations between the energy density, pressure and 4-velocity in both frames are given by $\rho=A^4(\varphi){\tilde \rho}$, $p=A^4(\varphi){\tilde p}$
and $u_{\mu}=A^{-1}(\varphi){\tilde u}_{\mu}$. More details on the relations between the two frames can be found in \cite{Yazadjiev2014,Doneva2013}.

Since we are interested in rapidly rotating neutron stars, we will concentrate on stationary and axisymmetric spacetimes as well as stationary and axisymmetric  fluid and scalar field configurations. Thus the metric can be written in the following general form
\begin{eqnarray}
ds_{*}^2 = - e^{2\nu}dt^2 + \rho^2 B^2 e^{-2\nu}(d\phi - \omega dt)^2 + e^{2\zeta - 2\nu}(d\rho^2 + dz^2),
\end{eqnarray}
where all the metric functions depend on $\rho$ and $z$ only. The reduced field equations are quite lengthy and we will not give them here, instead we refer the reader to \cite{Yazadjiev2015,Doneva2013}.

In addition to building and exploring equilibrium rapidly rotating neutron star models, we will study also universal relations for neutron stars in massive scalar-tensor theories and more precisely, we will focus on the relation between the normalized moment of inertia and quadrupole moment. These quantities are directly connected to the asymptotics of the metric functions given by
\begin{eqnarray}
&&\nu \approx - \frac{M}{r} + \left[ \frac{b}{3} + \frac{\nu_2}{M^3} P_{2}(\cos\theta)\right] \left( \frac{M}{r}\right)^3 , \label{ASMPT1}\\
&&B\approx 1 + b \left( \frac{M}{r}\right)^2, \label{ASMPT2}\\
&&\omega \approx \frac{2J}{r^3} , \label{ASMPT3}\\
&& \zeta\approx -\left\{\frac{1}{4}
+ \frac{1}{3}\left[b + \frac{1}{4}\right]\left[1 - 4P_{2}(\cos\theta)\right] \right\} \left( \frac{M}{r}\right)^2 , \label{ASMPT4}
\end{eqnarray}
where the quasi-isotropic coordinates $r$ and $\theta$ defined by $\rho=r\sin\theta , z=r\cos\theta$ are used for convenience and only terms up to $r^{-3}$ order are kept. Here $M$ and $J$ are the mass and the angular momentum, $b$, $\nu_2$ and $\varphi_{2}$ are constants and $P_{2}(\cos\theta)$ is the second Legendre polynomial. The scalar field decreases exponentially at infinity 
\begin{eqnarray}\label{eq:phi_asympt}
\varphi|_{ r\rightarrow\infty} \sim \frac{e^{-m_\varphi r}}{r}. 
\end{eqnarray}
Clearly the scalar charge, which is defined as the coefficient in front of the $1/r$ term in the scalar field expansion at infinity, is zero and does not play a role in the asymptotics of the metric functions contrary to the case of scalar-tensor theories with massless scalar field \cite{Doneva2014a}. Therefore, the quadrupole moment has the same form as in general relativity \cite{Pappas2012,Friedman2013}:
\begin{eqnarray}\label{eq:QuadrupoleMoment}
Q=- \nu_{2} - \frac{4}{3}\left[b + \frac{1}{4}\right]M^3.
\end{eqnarray}
The moment of inertia on the other hand, is defined in the standard way 
\begin{eqnarray}
I=\frac{J}{\Omega}.
\end{eqnarray}
where $\Omega$ is the angular frequency of the star. 

The formulas for the moment of inertia and the quadrupole moment are written in the Einstein frame but they are the same in the physical Jordan frame for the particular class of scalar-tensor theory we are using because of the exponential decay of the scalar field \cite{Doneva2014a,Doneva2015}.

In the next section where we present our numerical results we shall use the dimensionless parameter $m_{\varphi}\to m_{\varphi} R_{0}$  and the dimensionless moment of inertia $I\to I/M_{\odot}R^2_{0} $, where $M_{\odot}$ is the solar mass and $R_{0}=1.47664 \,{\rm km}$ is one half of the solar gravitational radius.

\section{Results}
In our studies we will employ three equations of state (EOS): APR, FPS and the zero temperature limit of the Shen EOS. The first one can be considered as modern realistic EOS which falls into the preferred range of masses and radii according to the observations. The Shen EOS is stiffer, the maximum mass of neutron stars reaches well above the two solar mass barrier and has larger radii. The FPS EOS is softer and it has maximum mass below two solar masses in pure general relativity. Nevertheless, it is useful in our studies since one of our goals is to check the EOS universality of the relations between the normalized moment of inertia and quadrupole moment which requires a broader set of EOS. Moreover, as we will show below, the presence of nontrivial massive scalar field significantly increases the maximum neutron star mass for a given EOS which can easily reconcile the theory with the observations.

A very important point we should comment are the observational constraints on the parameters of the theory, namely on the coupling parameters $\beta$ and the mass of the scalar field $m_\varphi$. In the massless case $\beta>-4.5$, according to the observations of neutron stars in close binaries \cite{Antoniadis13,Freire2012}. The presence of nonzero mass of the scalar field can drastically change this limit. More specifically, if the Compton wave-length of the scalar field $\lambda_\varphi=2\pi/m_{\varphi}$ is much smaller than the separation between the two stars in the binary system, then the emitted scalar gravitational radiation will be negligible and practically no constraints can be imposed on the parameter $\beta$.  If we further assume that the mass of the scalar field does not prevent the scalarizataion of the star we arrive at the following allowed range for the scalar-field mass:
\begin{equation} \label{eq:bounds_mphi}
10^{-16} {\rm eV} \lesssim m_\varphi \lesssim 10^{-9}{\rm eV}.
\end{equation}
This corresponds to roughly $10^{-6} \lesssim m_\varphi \lesssim 10$ in our dimensionless units. A mid-range of $m_\varphi$ values can also be excluded based on arguments connected with the superradiant instability of black holes, but since there are many uncertainties in these studies we will not impose further restrictions on $m_\varphi$ in the present paper and study the full parameter space given by eq. \eqref{eq:bounds_mphi}. The observational constrains are discussed in more details in \cite{Yazadjiev2016,Ramazanouglu2016}.

For values of $m_\varphi$ falling into the rage given by \eqref{eq:bounds_mphi}, $\beta$ is essentially unconstrained. In our paper we have chosen to work with a particular value of $\beta$, namely $\beta=-6$, for the following reason\footnote{Neutron stars with nontrivial scalar field, i.e. scalarized neutron stars, exist only for negative values of $\beta$.}. $\beta=-6$ gives already large deviations from pure general relativity even in the nonrotating case, that are further magnified for rapidly rotating models. That is why it can give us good qualitative intuition about the problem. Moreover, the deviations from pure general relativity at the Kepler (mass shedding limit) are already huge for $\beta=-6$. Our studies show that the results for smaller values of $\beta$ are qualitatively similar, only the magnitude of the deviations increases.

\begin{figure}[]
	\centering
	\includegraphics[width=0.45\textwidth]{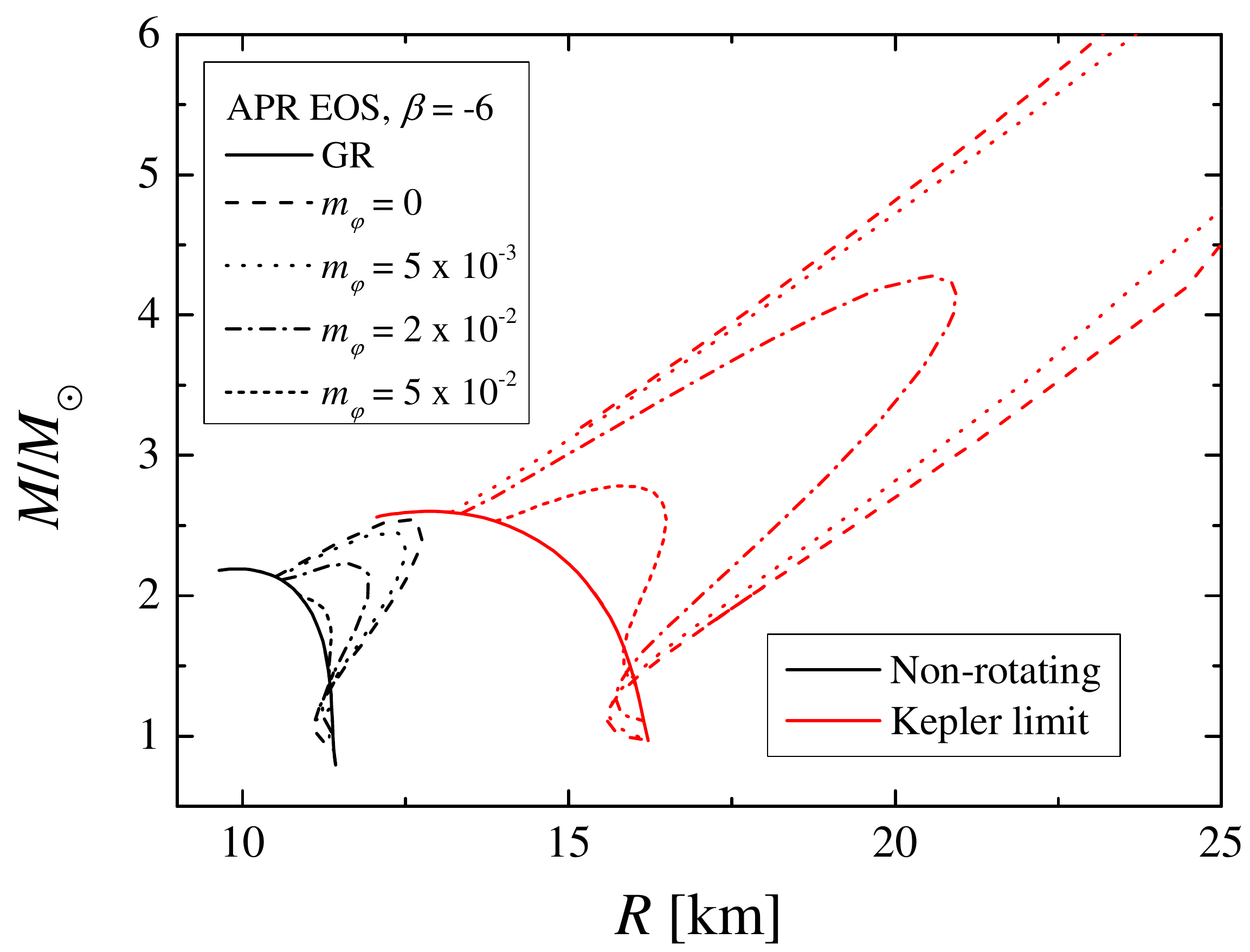}
	\includegraphics[width=0.45\textwidth]{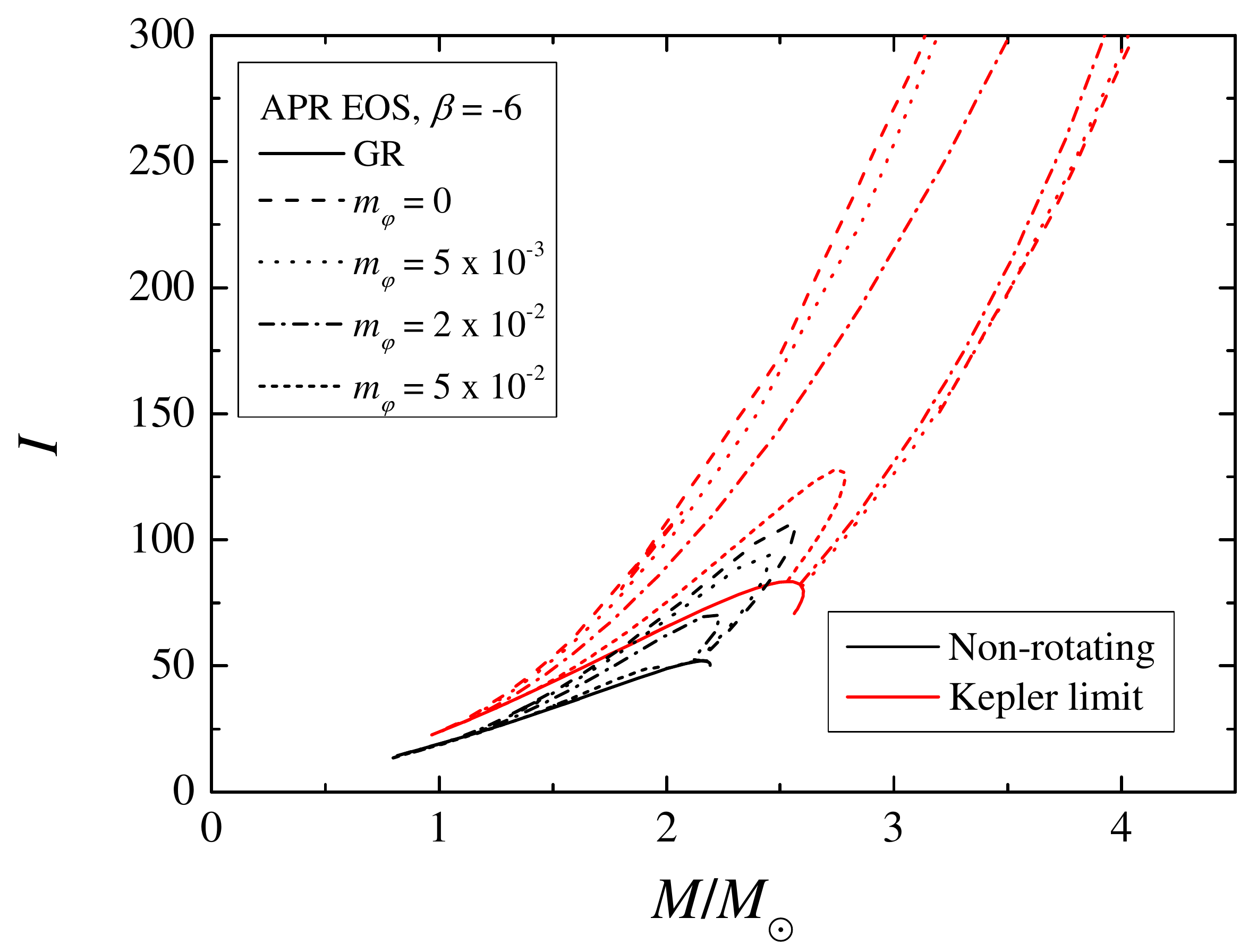}
	\caption{The mass as a function of the radius (left panel) and the moment of inertia as a function of the stellar mass (right panel).}
	\label{Fig:MR}
\end{figure}

In Fig. \ref{Fig:MR} the mass as a function of the radius is plotted in the left panel and the moment of inertia as a function of the mass in the right panel for the APR EOS. Sequences of models for $\beta=-6$ and several representative values of the scalar field mass are plotted in the nonrotating case and for models rotating at the Kepler limit. As one can see, the results for different values of $m_\varphi$ are situated between two limiting cases -- the general relativistic limit which loosely speaking corresponds to infinite mass of the scalar field\footnote{The Compton wave-length of the scalar field, and thus its range, is back proportional to the scalar field mass. As we increase the mass, the scalar field range decreases and in the limit of $m_\varphi\rightarrow \infty$ it is completely suppressed which is equivalent to the pure general relativistic case.} and the massless case $m_\varphi=0$. This is expected, since the scalar field mass actually suppresses the scalar field. Therefore, the effect of the scalar field mass is not that it produces on its own large deviations from Einstein's theory of gravity. Instead, the reason for the large differences is the  fact that the allowed range of $\beta$ is broadened significantly.

Strictly speaking the case of $m_\varphi=0$ in Fig. \ref{Fig:MR} is not allowed by the observations because in the massless case $\beta>-4.5$. However, our results show that the minimum possible value of $m_\varphi$ given by eq. \eqref{eq:bounds_mphi} leads to neutron star models that differ only marginally from the $m_\varphi=0$ case. That is why we can roughly assume that for a fixed value of $\beta$, the massless case $m_\varphi=0$ truly represents an upper limit on the deviations from pure general relativity. 

As one can see in Fig. \ref{Fig:MR}, the $\beta=-6$ case gives us already large deviations in the static case. The rapid rotation magnifies these deviations considerably and the  models rotating at the Kepler limit already differ drastically from pure general relativity. This effect is even more pronounces for the rotational properties of the star, such as the moment of inertia. Such a strong increase of the deviations from pure general relativity compared to the static case was also observed in other alternative theories of gravity \cite{Doneva2013,Yazadjiev2015}. As one can see, we have cut the rapidly rotating sequences up to a certain point in order to have a better visibility. The results show that for $\beta=-6$ in the limit of small scalar field masses the maximum mass in the Kepler limit is five times larger with respect to general relativity, compared to only 15\% difference in the static case. 
Of course, these numbers increase with the decrease of $\beta$. Such large deviations in the equilibrium properties of the stars should lead to observational manifestations that will allow us to put constraints on the massive scalar tensor theories that are tighter that the ones given by the binary pulsar observations. Such a study is underway.

\begin{figure}[]
	\centering
	\includegraphics[width=0.5\textwidth]{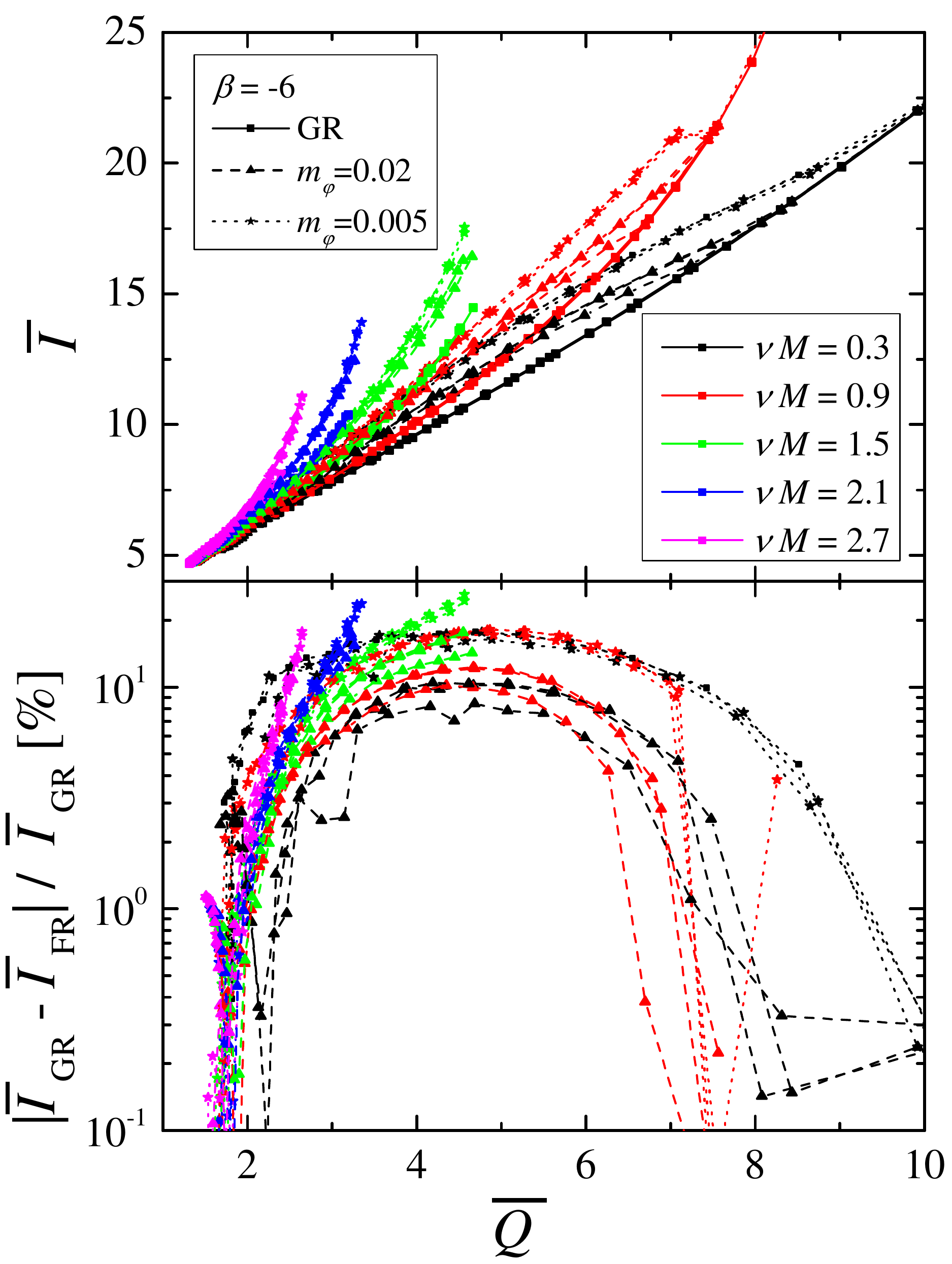}
	\caption{(upper panel) The ${\bar I}$ -- ${\bar Q}$ universal relations for sequences of models with fixed values of the normalized rotational frequency $\nu M$). (lower panel) The relative deviation from pure general relativity $\Delta I = \frac{\left|{\bar I}_{GR} - {\bar I}_{FR}\right|}{{\bar I}_{GR}}$.}
	\label{Fig:IQ}
\end{figure}

In Fig. \ref{Fig:IQ} the normalized moment of inertia as a function of the normalized quadrupole moment is shown. The normalization we use is the standard one: ${\bar I} \equiv I/M^3$ and ${\bar Q} \equiv Q/(M^3 \chi^2)$, where $\chi\equiv J/M^2$. We have used three EOS (APR, Shen and FPS) with different stiffness and thus they can be used as a good test of the EOS universality. Sequences with fixed rotational parameter $\nu M$ are plotted, where $\nu M$ has dimension of kHz times solar masses. The black lines with $\nu M=0.3$ correspond to the slow rotation limit. Again, we fixed $\beta=-6$ and examined two different values of $m_\varphi$, namely $m_\varphi=5 \times 10^{-3}$ and $m_\varphi=2 \times 10^{-2}$. The former case gives deviations close to the maximum ones \footnote{We did not plot the ${\bar I}-{\bar Q}$ relation for the $m_\varphi=0$ case for the following reason. When $m_\varphi=0$ the scalar charge is nonzero and it will contribute to the asymptotic of the metric and thus the quardupole moment. Therefore, eq. \eqref{eq:QuadrupoleMoment} is valid only for  $m_\varphi \neq 0$.}, and the latter one has intermediate deviations from general relativity.

We can make the following observations concerning the ${\bar I}-{\bar Q}$ relations. The relations are EOS independent up to a large extend for fixed values of $\nu M$. The EOS universality is a little bit worse for scalar tensor theories with nonzero  $m_\varphi$ compared to the general relativistic case, but the deviations still do not exceed a few percents. The most important observation though, is the fact that the differences with pure general relativity can be large for $\beta=-6$, especially for small scalar field masses, reaching above $20\%$ as the lower panel of Fig. \ref{Fig:IQ} shows. Our investigations show that these differences increase further with the decrease of $\beta$ that can be used to impose constraints on the parameters of the theory as suggested in \cite{Yagi2013,Yagi2013a}. 

The $I$-Love-$Q$ relations in most of the studied alternative theories of gravity are almost indistinguishable from pure general relativity. The reason is that the normalization used in the original  $I$-Love-$Q$ relation seems to be good enough not only to take away the EOS dependence, but also the dependence on the theory of gravity up to a large extend\footnote{Only the normalized relations are indistinguishable from pure general relativity, the unnormalized quantities can deviate significantly.}. The known exceptions are the dynamical Chern-Simons gravity \cite{Yagi2013,Yagi2013a} and the $f(R)$ theories of gravity \cite{Doneva2015}. As we have shown, the massive scalar-tensor theories lead also to large differences with the pure general relativistic case for values of the parameters that are in agreement with the observations. As a matter of fact, $f(R)$ theories are mathematically equivalent to a special class of scalar-tensor theories with nonzero potential for the scalar field. That is why taking into account that it was shown in \cite{Doneva2015} that $f(R)$ theories lead to significant differences compared with pure Einstein's theory, the results in the present paper are not surprising. What is interesting here, is that these differences can be very large for small values of $\beta$ and small scalar field masses.

\section{Conclusions}
In the present paper we have studied neutron stars in scalar-tensor theories with massive scalar field. We focused on a particular form of the coupling function which leads to interesting strong field effects such as scalarization and nonuniqueness of the solutions, but qualitatively similar results and large deviations from pure general relativity are expected for other choices of the coupling function as well. 

We have studied both equilibrium properties and universal (EOS independent) relations for rapidly rotating models with frequencies reaching the Kepler limit. The observed deviations from pure general relativity can be very large and the rapid rotation increases the differences several times especially for some of the rotational properties of the star, such as the moment of inertia. With the increase of the scalar field mass the deviations from pure Einstein's theory decrease. Therefore, the presence of scalar field mass alone can not serve directly as a mechanism for magnifying the differences. The reason for the large deviations lies in the fact, that the mass of the scalar field exponentially suppresses the scalar field at distances of the order of its Compton wave-length which can reconcile theory with the observations for a much broader range of parameters compared to the massless case. 

As far as EOS independent relations are concerned, we have focused on the relations between the normalized moment of inertia and quadrupole moment both for slowly and rapidly rotating models. Our results show that the deviations from pure general relativity can be large unlike most of the alternative theories of gravity considered in the literature, which can help us test the massive scalar-tensor theories of gravity. The normalized $\bar I$-$\bar Q$ relations are quite EOS independent for fixed values of the normalized rotational parameter, even though the deviations from EOS universality are generally larger than the pure general relativistic case.

At the end, we would like to further comment on the possible observational constraints we can impose on the massive scalar-tensor theories. As we have already discussed, the binary pulsar observations and the related gravitational wave emission can impose constraints only on the mass of the scalar field, i.e. the Compton wave-length should be less than the orbital separation between the two stars. After this requirement is satisfied we are left practically without any tight bounds on the coupling parameters. As shown in the present paper, the equilibrium properties and the universal relations for neutron stars in massive scalar tensor theories can differ dramatically from pure general relativity. This will inevitably lead to strong observational manifestations. Therefore, we will be able to set tight constraints on the parameters of the theory using the astrophysical observations. Such a study is underway.

\section*{Acknowledgements}
DD would like to thank the European Social Fund, the Ministry of Science, Research and the Arts Baden-W\"urttemberg and Baden Württemberg Foundation for the support. The support by the Bulgarian NSF Grant DFNI T02/6, Sofia University Research Fund under Grant 193/2016, ``New-CompStar'' COST Action MP1304 and ``CANTATA'' COST Action CA15117 is also  gratefully acknowledged.


\bibliography{references}

\end{document}